\global\let\ifmypprint\iffalse
\def\mypprint{\global\let\ifmypprint\iftrue}
\global\let\iftorefs\iffalse
\def\torefs{\global\let\iftorefs\iftrue}
\global\let\dofloatfig\iffalse
\def\floatthefig{\let\dofloatfig\iftrue}

\mypprint 
\ifmypprint
\documentstyle[twocolumn,prl,aps]{revtex}
\floatthefig
\else

  \iftorefs
    \catcode`\@=11
    
    \def\figure{\let\@capwidth\columnwidth\@float{figure}}
    \let\endfigure\end@float
    \@namedef{figure*}{\let\@capwidth\textwidth\@dblfloat{figure}}
    \@namedef{endfigure*}{\end@dblfloat}
    \catcode`\@=12
     \floatthefig
  \fi
\fi

\input epsf.tex

\begin{document}
\twocolumn[\hsize\textwidth\columnwidth\hsize\csname @twocolumnfalse\endcsname
\title{Traveling-Wave Chemotaxis}
\author{Raymond~E. Goldstein\cite{regemail}}
\address{Department of Physics, Joseph Henry Laboratories,
Princeton University, Princeton, NJ 08544}
\date{\today}
\maketitle
\begin{abstract}
A simple model is studied for the chemotactic movement of biological cells
in the presence of a periodic chemical wave.   It incorporates the feature of
{\it adaptation} that may play an important role in
allowing for ``rectified" chemotaxis: motion opposite the direction of
wave propagation.  The conditions under which such rectification occurs
are elucidated in terms of the form and speed of the chemical wave,
the velocity of chemotaxis, and the time scale for adaptation. An experimental
test of the adaptation dynamics is proposed.

\end{abstract}

\vskip2pc]

Many biological processes involve
{\it chemotaxis}, cellular motion in response to a chemical stimulus.
Often, the chemo-attractant propagates through a set of cells as traveling
waves \cite{chemotaxis_general,oscillations}, as in a case of
longstanding interest: the emergence of a
multicellular structure from colonies of the eukaryotic
microorganism {\it Dictyostelium discoideum} ({\it Dd})\cite{Bonner}.
In controlled experiments, a monolayer
with $10^5-10^6$ cells/cm$^2$ on the surface of agar
begins within several hours after nutrient deprivation to support
waves of cyclic adenosine monophosphate (cAMP) triggered by spontaneous
release of cAMP from a small subpopulation of cells.  These target or
rotating spiral waves (Fig. 1), whose fronts appear as bands under
dark-field visualization through their effects on cell shape \cite{Devreotes},
induce chemotaxis toward their centers, followed by complex
multicellular morphogenesis.

Chemical waves in excitable media such as {\it Dd}
are quite thoroughly explored \cite{Meron,Tyson}, but their
coupling to cell density through chemotaxis is far less well-understood,
although of longstanding interest \cite{Keller_Segel,Levine,Vasiev}.
As emphasized recently \cite{Wessels}, and illustrated in
Fig. 1a \cite{Lee}, chemotaxis driven by traveling waves is quite intriguing.
A cell in the position indicated by the arrow experiences a progression of
leftward-moving wavefronts as the nearby spiral rotates outward.
In seeking higher levels of cAMP, the cell
would move first rightward into each advancing wave, then leftward after the
peak has passed (Fig. 1b).
In the simplest model of chemotaxis, the
cell velocity is proportional to the local chemical gradient, and
it has been argued \cite{Wessels} (but not proven theoretically) that
the {\it net} cellular motion would be in the same direction as the wave:
i.e. ``advection" away from the center, rather than
the observed motion towards the spiral core.  Tracking studies of
cells \cite{Wessels} suggest a resolution to this by noting that
as the cells experience the rising cAMP level of the approaching wave,
their chemotactic response diminishes, leaving them less responsive to
the trailing edge, but their response recovers in time for the next front.
They thus {\it rectify} the traveling waves, with net motion
opposite that of the wave.

In an effort to understand the underlying mechanism of this process,
we study here a very simple model for ``adaptive"
traveling-wave chemotaxis and suggest experiments
to test its predictions for the conditions under which rectification
occurs.  This model is closely related to, but considerably
simpler than one introduced recently in important work by H\"ofer,
{\it et al.} \cite{Hofer}, who demonstrated by
numerical computations that a process of adaptation could
lead to rectified motion.  A number of important aspects of this problem
become clear with these simplifications, particularly in the experimentally
relevant limit of
chemotactic velocities small compared to the wave speed.
First, in this limit an elementary proof is given of the heuristic argument
\cite{Wessels} that nonadaptive chemotaxis will not produce rectified
motion.  Second, it is shown that rectified motion
requires only two main
\dofloatfig
\begin{figure}
\epsfxsize=2.8 truein
\centerline{\epsffile{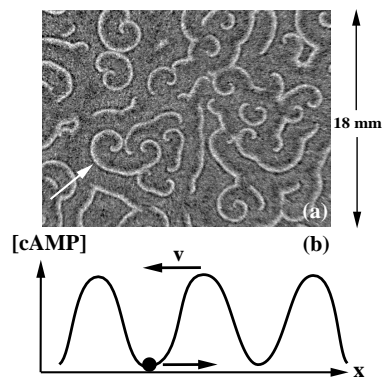}}
\smallskip
\caption[]{(a) Dark-field image of spiral waves in {\it Dictyostelium
discoideum}
\protect{\cite{Lee}}.  Wavefronts of cAMP appear as dark bands.
A cell in the position indicated
experiences a periodic train of cAMP waves, shown schematically in (b).
Net chemotactic motion occurs toward the spiral core, opposite the direction of
wave propagation.}
\label{fig_1}
\end{figure}
\fi
\noindent ingredients: (i) a single characteristic time for adaptation,
and (ii) a response function that decreases with concentration.

Third, the net chemotactic flux is
shown to have a thermodynamic
analogy in being proportional to the area enclosed by certain limit cycles
exhibited in the response-concentration plane.  Fourth, an
analytical calculation confirms the intuitive notion that rectification
is greatest when the adaptation time is comparable to the wave period
(as seen in experiment \cite{Wessels}), although
there can be a delicate interplay between the competing processes of
rectification and advection.  Finally, an experimental test of these
results is suggested.

Consider a one-dimensional set of noninteracting cells
at density $\rho$ responding to a periodic chemical concentration wave
$c(x,t)\equiv c(x+vt)$ with wavelength $\lambda$ and velocity $v$.
Typically, $\lambda \sim 0.1-0.5$ cm  (Fig. 1), and $v\sim 10^2-10^3$
$\mu$m/min.
We leave aside the complex dynamics of wave
production and its connection to the cell density
\cite{Cohen,Martiel_Goldbetter,Monk}.
The model is formulated at the level of the coordinate $x(t)$ of a cell,
and for the present purposes is deterministic, as the random motions of
the cells during one wave period are small on the scale of the wavelength
$\lambda$. Deterministic chemotaxis arising from chemical gradients is
described by the overdamped dynamics
\begin{equation}
{dx\over dt}= r {d\over dx}c(x+vt)~.
\label{first_eom}
\end{equation}
The {\it chemotactic response coefficient}
$r$ measures the
strength of chemotaxis, with cells migrating
to high values of $c$ when $r>0$.  When $r$ responds to $c$ we have
``adaptive chemotaxis"; otherwise the motion is nonadaptive.
The flux of cells $\langle J\rangle$ ($=(1/\lambda)\int_0^{\lambda}dz J$)
averaged over one wave period
is found by solving (\ref{first_eom}) in the moving frame $z=x+vt$,
with $dz/[r(dc/dz)+v]=dt$,
and transforming back (see also \cite{Faucheux}),
\begin{equation}
{\cal J}\equiv \langle J\rangle/\rho v=
\left<\left[1+r\left(dc/dz\right)/v\right]^{-1}\right>^{-1}-1~.
\label{flux_det}
\end{equation}

It is known from experiment that the typical chemotactic velocity
$r dc/dz$ in {\it Dd} is at least an order of magnitude lower than
the wave speed $v$ \cite{chemotaxis_general,Hofer}, so
we expand ${\cal J}$ in powers of $v^{-1}$,
\begin{equation}
{\cal J} \simeq {1\over v}\left<r{dc\over dz}\right>
- {1\over v^2}
\left[\left<\left(r{dc\over dz}\right)^2\right>
- \left<r{dc\over dz} \right>^2 \right]~.
\label{flux_det_large_v}
\end{equation}
In nonadaptive chemotaxis, $r$ is constant, so the first term
vanishes by the periodicity of $c(z)$.  The first nonvanishing contribution
to ${\cal J}$ is in the direction of the wave propagation, {\it independent of
the form of
$c(z)$} since the variance of $rdc/dz$ is manifestly positive.
This confirms the
heuristic argument of Wessels, {\it et al.} \cite{Wessels}.
The physical basis for this was emphasized in the context of
Brownian particles forced by moving optical traps \cite{Faucheux}:
Particles migrating into the advancing wave experience the leading edge
(and hence chemotax) for a shorter time than they do the trailing edge,
since their velocity relative to the wave is greater in the former case
than in the latter. One step forward, two steps back, so to speak.

A very simple adaptive chemotaxis model has two ingredients:
(i) an equilibrium ``adaptation function"
$f(c)$ that is a {\it decreasing} function of $c$,
and (ii) a single time constant $\tau$ for the relaxation of $r$ toward $f$
\cite{note},
\begin{equation}
\tau {dr\over dt}=f(c)-r~.
\label{r_dynamics}
\end{equation}
A cell having experienced a concentration
$c$ for times much longer than $\tau$ will have a response coefficient
$r=f(c)$ when next presented with a gradient: low when $c$ is high, and vice
versa.
As $c$ changes with time
the response will attempt to equilibrate to $f(c)$, but will
lag behind when $c$ changes on time scales shorter
than $\tau$.  One can think of this lag as a memory or inertial effect,
and it provides a means of rectification.  We expect the adaptation time
to be comparable to the refractory period of cAMP signaling.  Eq.
\ref{r_dynamics} is one member of a FitzHugh-Nagumo model that has
been studied in related work on chemotaxis \cite{Vasiev}.

Let us introduce the dimensionless coordinate $Z=kz$, time $T=\omega t$,
concentration $C=c/c_0$, response coefficient $R=r/f_0$, and adaptation
function $F=f/f_0$, where $k=2\pi/\lambda$, $\omega=kv$,
$c_0$ is the peak wave concentration, and $f_0 \equiv f(c=0)$.
The rescaled dynamics are
\begin{mathletters}
\label{rescaled_dyn}
\begin{eqnarray}
{dZ\over dT} &=& S R {dC\over dZ} +1 \\
\Omega{dR\over dT} &=& F(C)-R~,
\label{rescaled_pair}
\end{eqnarray}
\end{mathletters}
with two dimensionless parameters,
\begin{equation}
\Omega =\omega\tau~, \ \ S=f_0c_0k/v~.
\label{S_define}
\end{equation}
The quantity $\Omega$ measures the relaxation time in units of the
wave period, while $S$ is the ratio
of a characteristic chemotactic speed to the wave speed.

The expansion of the average flux in
Eq. (\ref{flux_det_large_v}) now appears as an expansion in $S$, with
the ${\cal O}(S)$ term possibly rectifying, and those
of order $S^2$ always advective.
For a given value of $S$, we expect three regimes of $\Omega$: (i) $\Omega\gg
1$,
the nonadaptive case already discussed, (ii) $\Omega\simeq 1$, where
rectification
may occur, and (iii) $\Omega \ll 1$,
with instantaneous adaptation.  In region (iii), the
response tracks the concentration
precisely, and the leading and trailing sides of the wave
are not distinguished, so the flux is positive and
given analytically by Eq. (\ref{flux_det}) with $r$ replaced by $f(c)$.
It is smaller than in (i) since $f(c)$ is smaller for high $c$.
Rectification may occur in region (ii), where the
down-regulation of the response
triggered by the advancing wave has not fully recovered by the time the
trailing edge is encountered.
\dofloatfig
\begin{figure}
\epsfxsize=2.7 truein
\centerline{\epsffile{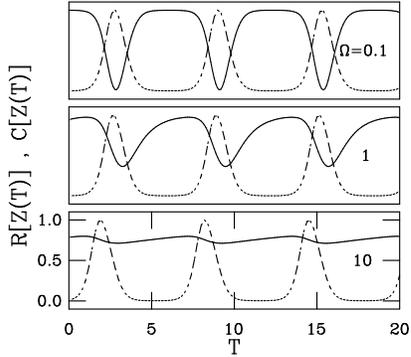}}
\smallskip
\caption[]{Relation between morphogen concentration and chemotactic response.
Panels show the imposed chemical wave $C(Z(T))$ (dashed) and the
response coefficient $R(Z(T))$ (solid) for $S=0.2$ and various scaled
frequencies $\Omega$.}
\label{fig_2}
\end{figure}
\fi

Figure \ref{fig_2} shows for
different values of $\Omega$ the concentration $C$ and response coefficient $R$
as functions of time along the trajectory of a moving cell obtained by
numerical integration of (\ref{rescaled_dyn}).
An arbitrary initial condition decays in a time of order $\Omega$
into the steady patterns shown.
To simulate the sharply-peaked traveling waves seen in experiment, we
chose $C(Z)=\exp\left[\beta\left(\sin(Z)-1.0\right)\right]$, with
$\beta=3.0$.  The response function is the simplest: a linearly
decreasing function of $C$: $F(C)=1-C$.
For $\Omega=0.1$ (close to instantaneous adaptation) we see $R$ almost
precisely
anticorrelated with $C$, whereas for
$\Omega\simeq 1$ the asymmetric response between leading and trailing edges is
quite
apparent. For $\Omega=10$ the response settles to a nearly constant value
determined by the
mean value of the concentration over one wave period.

\begin{figure}
\epsfxsize=2.7 truein
\centerline{\epsffile{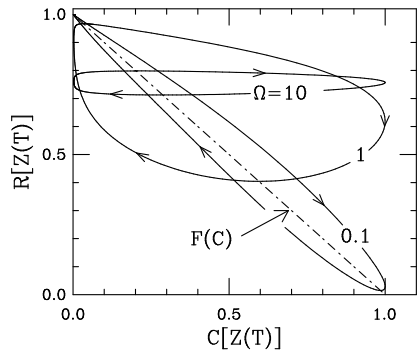}}
\smallskip
\caption[]{Phase portrait of traveling-wave chemotaxis.  Data in Fig.
\protect{\ref{fig_2}} are replotted in the $R-C$ plane.  The
equilibrium function $F(C)$ is shown dashed.  Arrows indicate
sense of traversal with increasing time.}
\label{fig_3}
\end{figure}

\begin{figure}
\epsfxsize=2.7 truein
\centerline{\epsffile{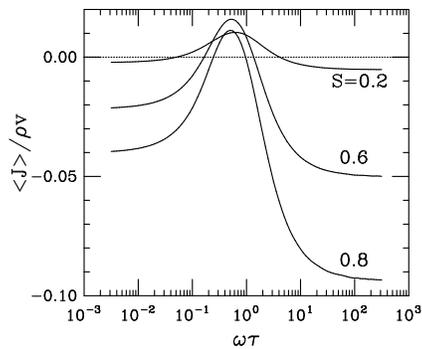}}
\smallskip
\caption[]{Flux as a function of relaxation time and
chemotactic velocity.
At low $S$, rectification occurs over a broad range of $\Omega$, while at
higher $S$ advection is dominant except over a narrow range of
$\Omega\simeq 1$.}
\label{fig_4}
\end{figure}

The extent to which the response is ``out of equilibrium" is
seen with limit cycles in the $R-C$ phase plane shown in Fig. \ref{fig_3},
with position $Z$ as a parameter.
For small $\Omega$ the cycle hugs the equilibrium curve $F(C)$,
while when $\Omega \gg 1$ it is a narrow loop encircling a
horizontal line of constant $R$.
In the rectifying regime ($\Omega=1$) the cycle lies very far from
equilibrium, forming a large closed loop.
Using the high-velocity result in Eq. (\ref{flux_det_large_v}),
we may express the chemotactic flux directly in terms of the area enclosed by
this loop ${\cal C}$,
\begin{equation}
{\cal J} \simeq {S\over 2\pi} \int_0^{2\pi}\! dZ R {dC\over dZ}
={S\over 2\pi} \oint_{{\cal C}} R dC~,
\label{loop_integral}
\end{equation}
in much the same way as we associate mechanical work with loops in the
pressure-volume plane.  With the sense of traversal of the loops
shown in Fig. \ref{fig_3} this area is positive, and hence rectifying.
With these results, we obtain the chemotactic flux shown in
Fig. \ref{fig_4}, highlighting the window around $\Omega=1$
within which rectification occurs.
For $S$ small rectification occurs over almost the entire range of
$\Omega$, since
advection is negligible.  But for larger
\dofloatfig
\fi
\noindent $S$ advection dominates at the
extremes of $\Omega$, and rectification occurs in a very narrow window
near $\Omega=1$.
The qualitative behavior seen in Figs. \ref{fig_2}-\ref{fig_4} is unchanged
by the inclusion of more complicated nonlinear adaptation functions more
faithful to
the biochemistry of receptor binding \cite{Hofer}.
\dofloatfig
\fi

Insight into the quantitative behavior of the model may be obtained by an
analytic calculation in the limit of large wave speeds (small $S$)
\cite{Hofer}.
We assume an expansion of the position $Z$ and response coefficient
$R$ in powers of $S$:
$Z(T) = Z^{(0)}(T)+S Z^{(1)}(T) + \cdots$ and
$R(T) = R^{(0)}(T)+S R^{(1)}(T) + \cdots$.
At order $S^0$ in this moving coordinate system we obtain $dZ^{(0)}/dT=1$, so
$Z^{(0)}(T)=T$.  At order $S$
we obtain the equation of motion of a particle due to a time-dependent force
\begin{mathletters}
\label{high_vel}
\begin{equation}
{dZ^{(1)}\over dT}= R^{(0)}(T){dC(T)\over dT}~,
\label{first_order_pair}
\end{equation}
\begin{equation}
R^{(0)}(T)={1\over \Omega}\int^T\! dT' {\rm e}^{-(T-T')/\Omega}F(C(T'))~.
\label{R_solved}
\end{equation}
\end{mathletters}

We continue with the simple model $F(C)=1-C$.  If $C(Z)$ has the
Fourier expansion $C(Z)=\bar C + \sum_{n=1}^{\infty}\left[ \alpha_n
\exp(i n Z)+c.c.\right]$,  then after the transients the modes of $R$
are shifted in phase from the wave,
\begin{equation}
R^{(0)}(T)=1-\bar C
-\sum_{n=1}^{\infty}{\alpha_n {\rm e}^{i(n T-\theta_n)}+ c.c.\over
\sqrt{1+(n \Omega)^2}}~.
\label{response_fast}
\end{equation}
The phase shifts $\theta_n$ satisfy
$\tan\theta_n=n\Omega$,
and thus are nearly zero for instantaneous adaptation and tend to
$\pi/2$ as $\Omega\to \infty$, more rapidly with increasing mode number $n$.
To compute the flux in this small-$S$ limit, we find the mean
value of $R^{(0)}(T)dC/dZ$ over one period,
\begin{equation}
{\cal J} \sim {S\over 2\pi}\sum_{n=1}^{\infty}n \sin(2\theta_n)
\left\vert \alpha_n\right\vert^2~,
\label{analytical_flux}
\end{equation}
where $\sin(2\theta_n)= 2 n \Omega/( 1+ (n \Omega)^2)$.
The generic behavior of this sum is seen from its first term, which
vanishes as $\Omega\to 0$ and $\Omega\to \infty$, and peaks
at $\Omega=1$ to give maximum forward flux.  When the wave has dominant
spectral
weight in a higher mode $n$, the peak in flux will occur for $\Omega\sim 1/n$,
as in Fig. 4.
Once we go beyond the limit of vanishing $S$, the
negative contributions to the flux (\ref{flux_det_large_v}) compete with the
rectifying part to produce the resonance-like behavior seen in Fig.
\ref{fig_4}.

We see that rectified chemotactic motion
requires only two simple ingredients: a single time scale for
adaptation, and an adaptation function
that decreases with concentration.  Adaptive
phenomena are found in many biological systems besides {\it D. discoideum},
including those of bacterial chemotaxis \cite{Shapiro}.
While this problem may appear similar to ones of
directional transport studied recently in the context of electrophoresis
\cite{Ajdari}, there is a fundamental distinction;
the rectified motion arises here through processes internal
to the particles, not through stochasticity.
Indeed, the random motions of cells are expected to decrease the
efficiency of rectification.
Finally, these results suggest experiments to probe the competition between
advective and adaptive chemotaxis (Fig. \ref{fig_4})
with artificially-produced chemical
waves of controllable shape and velocity \cite{austin}, complementary to recent
experiments
with fixed gradients \cite{Fisher} and time-varying uniform concentrations
\cite{Wessels}.  Observation of the predicted advection with a wave
whose frequency is either very small or very large compared to the natural
signalling frequency
would provide important evidence that adaptation associated with an
internal time scale is the operative mechanism in rectified chemotaxis.

I am indebted to J.T. Bonner, E.C. Cox, K.J. Lee, R.H. Austin,
and P. Nelson for many discussions, to P. Holmes,
and C.H. Wiggins for suggestions, and to A. Goriely
for pointing out the work of Hofer, {\it et al.}.
This work was supported by NSF PFF grant
DMR 93-50227, and the A.P. Sloan Foundation.

\end{document}
\end